\documentclass[12pt,english]{article}
\usepackage[latin1]{inputenc}
\usepackage{amssymb}
\usepackage{marvosym}
\makeatletter

\newif\iffigs
  \figsfalse

\usepackage{latexsym}

\usepackage{colordvi}

\iffigs

    \input{epsf}
\else
\fi
\def\tilde{\widetilde}

\newsavebox{\uuunit}
\sbox{\uuunit}
                             {\setlength{\unitlength}{0.825em}
                                      \begin{picture}(0.6,0.7)
                                                                  \thinlines
                                                                  \put(0,0){\line(1,0){0.5}}
                                                                  \put(0.15,0){\line(0,1){0.7}}
                                                                  \put(0.35,0){\line(0,1){0.8}}
                                                                 \multiput(0.3,0.8)(-0.04,-0.02){10}{\rule{0.5pt}{0.5pt}}
                                      \end {picture}}

\makeatletter \@addtoreset{equation}{section} \makeatother

\def\dim{\mathop{\rm dim}\nolimits}

\newcommand{\GL}{\mathop{\rm GL}}
\newcommand{\Hom}{\mathop{\rm Hom}}

\def\bfone{\relax{\rm 1\kern-.35em 1}}

\def\tilde{\widetilde}
\def\hat{\widehat}
\def\n010{N^{0,1,0}}
\def\tn010{\tilde{N}^{0,1,0}}
\def\det{\mathrm{det}}

\usepackage{ifpdf}
\ifpdf
   \pdfoutput=1
   \pdftrue
  \usepackage[pdftex]{hyperref}
 \else
   \usepackage{cite}
\fi

\newtheorem{theorem}{Theorem}

\usepackage{babel}
\makeatother
\begin{document}
\begin{titlepage}

\begin{flushright}
 DISTA-2007 
 \par\end{flushright}

\vskip 1.5cm

\begin{center}\textbf{\LARGE Balanced Superprojective Varieties} \textbf{ }\\
 \textbf{\vfill{}
} \textbf{\large R. Catenacci$^{~\small\mbox\Bicycle, \mbox\Mobilefone}$, M. Debernardi$^{~\small\mbox\Bicycle}$, } \\
 \textbf{\large P.A. Grassi$^{~\small\mbox\Bicycle, \mbox\Mobilefone, \mbox\Coffeecup}$, and 
 D. Matessi$^{~\small\mbox\Bicycle}$
} 
{
\vfill{}
{\Bicycle  ~{ DISTA, Universit\`a del Piemonte Orientale,}}\\
{ Via Bellini 25/G, Alessandria, I-15100, Italy,}  \\
{\Mobilefone  ~INFN - Sezione di Torino-Gruppo collegato di Alessandria}\\
{\Coffeecup ~ { Centro Studi e Ricerche E. Fermi, Compendio Viminale,
I-00184, Roma, Italy.}
 }}
 \par\end{center}

\vfill{}

\begin{abstract}
We first review the definition of superprojective spaces from 
the functor-of-points perspective. We derive the relation between superprojective 
spaces and supercosets in the framework of the theory of sheaves. As 
an application of the geometry of superprojective spaces, we extend Donaldson's definition of balanced manifolds 
to supermanifolds and we derive the new conditions of a balanced supermanifold. 
We apply the construction to superpoints viewed as submanifolds of superprojective 
spaces.  We conclude with a list of open issues and interesting problems that can be 
addressed in the present context.

\end{abstract}
\vfill{}
 \vspace{1.5cm}
 \vspace{2mm}
 \vfill{}
 \hrule width 3.cm 
 {\small catenacc@mfn.unipmn.it, marcod@mfn.unipmn.i, pgrassi@cern.ch, and matessi@mfn.unipmn.it}
 \end{titlepage}


\section{Introduction}

Supermanifolds are rather well-known in supersymmetric theories and
in string theory. They provide a very natural ground to understand the supersymmetry and supergravity from a geometric point of view. Indeed, a supermanifold contains the anticommuting coordinates which are needed to
construct the superfields whose natural environment is the graded algebras
\cite{wess-bagger,manin}. However, the best way to understand the supermanifold is using the theory of sheaves \cite{manin,Bruzzo}.
In the present notes we review this approach and its usefulness in theoretical physics and in particular in the last developments (twistor string theory \cite{Witten:2003nn} and pure spinor string theory \cite{Berkovits:2000fe}).

In the case of twistor string theory, the
target space is  indeed the supermanifold $\mathbb{CP}^{(3|4)}$
which can be described in two ways: as a supercoset of the supergroup
$PSU(4|4)/SU(3|4)$ or as a quotient of the quadratic 
hypersurface in the superspace $\mathbb{C}^{(4|4)}$ given by
\begin{equation}\label{oneA}
\sum_{\alpha \dot\alpha}  |Z^{\alpha\dot\alpha}|^{2} + \sum_{A} \bar \psi_{A}\psi^{A} = 1
\end{equation}
where $(Z^{\alpha\dot\alpha},\psi^{A})$ are the supertwistor coordinates.
Obviously, this equation needs a clarification: the commuting coordinates
$Z^{\alpha\dot\alpha}$ cannot be numbers for the above equation
to have a non-trivial meaning. One way to  interpret the above equation is using
the sheaf point of view where $Z^{\alpha\dot\alpha}, \psi^{A}$ are the
generators of a sheaf of supercommuting algebra over open
sets on $\mathbb{CP}^{3}$. In this way, the supermanifold can be viewed as
\begin{equation}
(\mathbb{CP}^{3}, {\cal O}_{\mathbb{CP}^{3}}(Z^{\alpha\dot\alpha}, \psi^{A}))
\end{equation}
and the equation (\ref{oneA}) makes sense (see also \cite{Schwarz:1995ak}). The
second way is using the {\it functor of point}. This is a functor
between the category of sets and the category of supermanifolds and,
as is well explained in \cite{Varadarajan:note} and the forthcoming
sections, it assigns a point in a supermanifold in terms of a
set of coordinates. The easiest way to realize the functor of point
is to map a superspace into a supermanifold and describe the
latter in terms of points identified by morphisms. 
Concretely, this
amount to choose a graded algebra with $N$ generators and represent
the generators of the sheaf $ {\cal O}_{\mathbb{CP}^{3}}(Z^{\alpha\dot\alpha}, \psi^{A})$ in terms of them. Then inserting this decomposition
in (\ref{oneA}), one gets a set of numerical equations for the coefficients
of the decomposition and they can be solved or studied by the conventional
means of algebraic geometry.

Of course the hypersurface (\ref{oneA}) is one example of
manifold that can be realized in terms of the generators
of ${\cal O}_{\mathbb{CP}^{3}}(Z^{\alpha\dot\alpha}, \psi^{A})$ and
that can be studied by means of the functor of points. Notice that
also from the supercoset point of view, the technique of the functor of point
gives us a representation of the supercoset in terms of the generators of
a sheaf. Indeed, by multiplying supermatrices (whose entries are
the generators of the sheaf) one finds that the entries cannot be numbers
and they have to be promoted to the generator of a sheaf. Therefore the
multiplication between matrices and the group multiplication
of a supergroup has to be understood as a morphism of a ringed space.
This point of view has been emphasized by Manin \cite{manin} and
recently by \cite{Varadarajan:note,Fioresi:2006gx}. 
We provide here a more elementary explanation of the role of functor-of-point in the
case of supergroup and supercosets. The purpose of this is to
use the functor-of-point to define the superprojective 
spaces (as $\mathbb{CP}^{(3|4)}$ above) and to prove the isomorphism with
the supercoset point-of-view as in the purely bosonic case.

In the second part of the paper, we develop two applications for superprojective spaces. Following the recent analysis of Donaldson \cite{dona} on balanced manifold, we extend his definition to supermanifolds.  
One ingredient is the definition of balanced submanifold of a projective space (for example a point or a line). For 
that we extend the integral equation given in \cite{dona} to an integration on the supermanifold. The definition of 
 the integral of a superform in a supermanifold is not an obvious extension since a regularization is needed. This can 
 be done using the projection forms as illustrated in \cite{Grassi:2004tv} and discussed in more detail 
in \cite{Berkovits:2006vi}. We briefly discuss this point in the text, but we refer to a forthcoming publication 
for a more detailed account \cite{mare}. 

After discussing the general theory, we provided a simple example 
of the embedding of $\mathbb{P}^{1|2}$ into the superprojective space $\mathbb{P}^{2m-1|2m}$ of sections 
$H^0(\mathbb{P}^{1|2}, L^{\otimes m})$ where $L^{\otimes m}$ is the m-power of a line bundle L over 
$\mathbb{P}^{1|2}$. In this case both the base manifold and the sections $\mathbb{P}^{2m-1|2m}$ are super-Calabi-Yau spaces (in the sense that they are super-K\"ahler spaces with vanishing Ricci tensor and an holomorphic 
top form $\Omega_{CY}$) and for those there is a natural measure for integrating superforms provided 
by $\Omega_{CY}\wedge \bar\Omega_{CY}$. It is shown that there are two types of conditions 
emerging from the extension of the Donaldson equations to the supermanifold case and therefore 
this restricts the number of supermanifolds that can be balanced subvarieties of superprojective spaces.  In generalizing the analysis of Donaldson we have taken into account the extension of the Kodaira embedding theorem discussed in \cite{lebrun}. 

The second application is to consider a set of points $\mathbb{C}^{0|N}$ immersed 
in the superprojective space $\mathbb{P}^{1|N}$ as a subvariety.  In this case we computed 
explicitly the general expression for 
the case $\mathbb{C}^{0|2}$ embedded into $\mathbb{P}^{1|2}$ and we found the condition for the balancing 
of a point. We found also how the supermanifold case generalizes the classical embedding 
condition and we argued 
how one can recover the classical balancing in addition to the requirements on 
the parameter of the superembeddings. We showed 
that this is tied to the choice of the integration measure for superforms. 

A concluding remark: we have not explored all possible implications of our extension neither we have 
discussed the relation with the stability of points in the sense of Geometric Invariant Theory (GIT) 
\cite{thomas} Nevertheless we have found rather interesting that some applications admit a non trivial 
generalization of the usual geometric setting. These results open new questions about the geometry of sheaves and 
their functor-of-point interpretation. 

The paper is organized as follows: in sec. 2 we define the supermanifolds form a sheaf theory point 
of view. We discuss the basic architecture and the set of morphisms. In sec. 3 we define superprojective spaces  and in sec. 4 we provide a functor-of-point interpretation Part of this material is summary 
of notes \cite{Varadarajan:note}. This allows us to use the local coordinates and 
to define the concept of a point in a supermanifold. 
In sec. 5, we study supergroups and superdeterminant (Berezinians) from the functor-of -point perspective needed to see the definition of superprojective space 
as supercosets of supergroups discussed in sec. 5.1.  In sec. 6, we extend the construction of Donaldson to 
supermanifold and we define balanced supermanifolds. Finally, in sec. 6.2 we discuss the balancing of points 
in superprojective spaces.

\section{Supermanifolds}
\label{defs}

\subsection{Definitions}

A \textbf{super-commutative} ring is a $\mathbb{Z}_2$-graded ring $A = A_0 \oplus A_1$ such
that if $i,j \in \mathbb{Z}_2$, then $a_i a_j \in A_{i+j}$ and $a_i a_j = (-1)^{i+j}a_ja_i$, where $a_k \in A_k$. Elements in
$A_0$ (resp. $A_1$) are called \textbf{even} (resp. \textbf{odd}).

A \textbf{super-space} is a super-ringed space such
that the stalks are local super-commutative rings (Manin-Varadarajan).
Since the odd elements are nilpotent, this reduces to require that
the even component reduces to a local commutative ring.

A \textbf{super-domain} $U^{p|q}$ is the super-ringed space $\left(U^{p},\mathcal{C}^{\infty p|q}\right)$,
where $U^{p}\subseteq\mathbb{R}^{p}$ is open and $\mathcal{C}^{\infty p|q}$
is the sheaf of super-commutative rings given by: 
\begin{equation}
V \mapsto \mathcal{C}^{\infty}\left(V\right)\left[\theta^{1},\theta^{2},...,\theta^{q}\right],\end{equation}
where $V \subseteq U^p$ is   and $\theta^{1},\theta^{2},...,\theta^{q} $ are
generators of a Grassmann algebra. The grading is the natural grading in even and odd elements. The notation is taken from \cite{Varadarajan:2004yz} and from the notes \cite{Varadarajan:note}.

Every element of $\mathcal{C}^{\infty p|q}\left(V\right)$ may be
written as $\sum_{I}f_{I}\theta^{I}$, where $I$ is a multi-index.
A \textbf{super-manifold} of dimension $p|q$ is a super-ringed space
locally isomorphic, as a ringed space, to $\mathbb{R}^{p|q}$. The coordinates $x_{i}$
of $\mathbb{R}^{p}$ are called the even coordinates (or bosonic), while the coordinates
$\theta^{j}$ are called the odd coordinates (or fermionic). We will denote by $\left(M,\mathcal{O}_{M}\right)$ the supermanifold whose underlying topological space is $M$ and whose sheaf of super-commutative rings is $\mathcal{O}_{M}$. 

To a section $s$ of $\mathcal{O}_{M}$ on an open set containing $x$ one may associate the 
\textbf{value} of $s$ in $x$ as the unique real number $s^{\sim}\left(x\right)$ such that 
$s-s^{\sim}\left(x\right)$ is not invertible on every neighborhood
of $x$. The sheaf of algebras $\mathcal{O}^{\sim}$, whose sections are the functions 
$s^{\sim}$, defines the structure of a differentiable manifold on $M$, called the
\textbf{reduced manifold} and denoted $M^{\sim}$.


\subsection{Morphisms.}

In order to understand the structure of supermanifolds it is useful to study their morphisms. Here we describe how a morphism of supermanifolds looks like locally. A \textbf{morphism} $\psi$ from $(X, \mathcal{O}_{X})$ to $(Y, \mathcal{O}_{Y})$ is given by a smooth map
map $\psi^{\sim}$ from $X^{\sim}$ to $Y^{\sim}$ together with a sheaf map:
\begin{equation}
\psi_{V}^{\ast}:\mathcal{O}_{Y}(V) \longrightarrow \mathcal{O}_{X}(\psi^{-1}(V)), 
\end{equation}
where $V$ is open in $Y$. The homomorphisms $\psi_{V}^{\ast}$ must commute
with the restrictions and they must be compatible with the super-ring
structure. Moreover they satisfy
\[ \psi_{V}^{\ast}(s)^{\sim} = s^{\sim} \circ \psi^{\sim}. \]

We illustrate this with an example taken from \cite{Varadarajan:note}. 
Given $M = \mathbb{R}^{1|2}$, we describe a morphism $\psi$ of $M$ into itself such that
$\psi^{\sim}$ is the identity. Let $\psi^{\ast}$ be the pull-back
map defined previously. We denote $\{ t,\theta^{1},\theta^{2}\}$ the coordinates on $M$, 
where $t$ can be interpreted both as the coordinate on $M^{\sim}=\mathbb{R}$ or as an even 
section of the sheaf. Since the sheaf map must be compatible with the $\mathbb{Z}_{2}-$grading, $\psi^{\ast} t$ is an even section and $(\psi^{\ast} t)^{\sim}=t$.
Then,
\[ 
\psi^{\ast}(t)=t+f(t)\theta^{1}\theta^{2}.
\]
Similarly,
\[ 
\psi^{\ast}(\theta^{j})=\, g_{j}(t)\theta^{1}\,+\, h_{j}(t)\theta^{2}.
\]
It is important to observe that this defines uniquely $\psi^{\ast}$ for sections of the 
form
\[ 
   a+b_{1}\theta^{1}+b_{2}\theta^{2}.
\]
where $a, b_{1}$ and $b_{2}$ are polynomials in $t$. It is therefore reasonable to expect 
that $\psi^{\ast}$ is uniquely defined. Let us take, for simplicity, the case where
\[ \psi^{\ast}(t)=t+\theta^{1}\theta^{2}, \]
and
\begin{equation} \label{psi:theta}
 \psi^{\ast}(\theta^{j})=\theta^{j}. 
\end{equation}
If $g$ is a smooth function of $t$ on an open set $U\subseteq\mathbb{R}$,
we want to define $\psi_{U}^{\ast}(g)$.

Let us expand $g(t+\theta^{1}\theta^{2})$ as a formal Taylor series:
\[ g(t+\theta^{1}\theta^{2})\,=\, g(t)\,+\, g'(t)\theta^{1}\theta^{2}. \]
The series does not continue because $(\theta^{1}\theta^{2})^{2}=0$. Then, we define
\[ \psi_{U}^{\ast}(g)\,=\, g(t)\,+\, g'(t)\theta^{1}\theta^{2}. \]
If
\[ g=g_{0}\,+\, g_{1}\theta^{1}\,+\, g_{2}\theta^{2}\,+\, g_{12}\theta^{1}\theta^{2}, \]
then we must define
\[ \psi_{U}^{\ast}(g)\,=\,\psi_{U}^{\ast}(g_{0})\,+\,\psi_{U}^{\ast}(g_{1})\theta^{1}\,+\,\psi_{U}^{\ast}(g_{2})\theta^{2}\,+\,\psi_{U}^{\ast}(g_{12})\theta^{1}\theta^{2}. \]
where we have used (\ref{psi:theta}).
The family $(\psi_{U}^{\ast})$ then defines a morphism between $\mathbb{R}^{1|2}$
and itself. This method can be extended to the general case. 

Let us recall some fundamental local properties of morphisms. A morphism
$\psi$ between two super-domains $U^{p|q}$ and $V^{r|s}$ is given
by a smooth map $\psi^{\sim}:U\rightarrow V$ and a homomorphism of
super-algebras \[
\psi^{\ast}:\mathcal{C}^{\infty\, r|s}(V)\rightarrow\mathcal{C}^{\infty\, p|q}(U).\]
It must satisfy the following properties:
\begin{itemize}
\item If $t=(t_{1},\ldots,t_{r})$ are coordinates on $V^{r}$, each component
$t_{j}$ can also be interpreted as a section of $\mathcal{C}^{\infty\, r|s}(V)$.
If $f_{i}=\psi^{\ast}(t_{i})$, then $f_{i}$ is an even element of
the algebra $\mathcal{C}^{\infty\, p|q}(U)$.
\item The smooth map $\psi^{\sim}:U\rightarrow V$
must be $\psi^{\sim}=(f_{1}^{\sim},\ldots,f_{r}^{\sim})$, where the $f_{i}^{\sim}$ are the 
values of the even elements above.
\item If $\theta_{j}$ is a generator of $\mathcal{C}^{\infty\, r|s}(V)$,
then $g_{j}=\psi^{\ast}(\theta_{j})$ is an odd element of the algebra
$\mathcal{C}^{\infty\, p|q}(U)$.
\end{itemize}
The following fundamental theorem (see for example \cite{Varadarajan:note}) gives a local 
characterization of morphisms:

\begin{theorem}{[}\textbf{Structure of morphisms}] \label{morphisms}
Suppose $\phi:U\rightarrow V$ is a smooth map and $f_{i},g_{j}$,
with $i=1,\ldots,r$, $j=1,\ldots,s$, are given elements of $\mathcal{C}^{\infty\, p|q}(U)$,
with $f_{i}$ even, $g_{j}$ odd and satisfying $\phi=(f_{1}^{\sim},\ldots,f_{r}^{\sim})$.
Then there exists a unique morphism $\psi:U^{p|q}\rightarrow V^{r|s}$
with $\psi^{\sim}=\phi$ and $\psi^{\ast}(t_{i})=f_{i}$ and $\psi^{\ast}(\theta_{j})=g_{j}$.
\end{theorem}
\medskip
\textbf{Remark.}  If $V$ is a vector bundle over a smooth manifold $M$, then we can 
form its \textbf{exterior bundle} $E = \Lambda^{max} V$. Let $\mathcal{O}(E)$ be the 
sheaf of sections of $E$. Then, locally on $M$, the sheaf is isomorphic to $U^{p|q}$ where 
$p=dim(M)$ and $q=rank(V)$. This is clearly true whenever $V$ is restricted to some open subset 
of $M$ over which it is trivial. Consequently, $(M, \mathcal{O}(E))$ is a super-manifold, 
denoted by $E^{\flat}$. Every super-manifold is locally isomorphic to a super-manifold of the 
form $E^{\flat}$. However we should note the important fact that $E^{\flat}$, as a 
supermanifold, has many more morphisms than the corresponding exterior bundle $E$, because of 
the possibility that the even and odd coordinates can be mixed under transformations.
This is well illustrated by the previous simple example. Another way to say the same thing 
is that there are less morphisms which preserve the bundle structure than morphisms 
which preserve the super-manifold structure.

\subsection{Local charts on supermanifolds} \label{smfld}
We describe how supermanifolds can be constructed by patching local charts.
Let $X=\bigcup_{i}X_{i}$ be a topological space, with $\{ X_{i}\}$
open, and let $\mathcal{O}_{i}$ be a sheaf of rings on $X_{i}$,
for each $i$. We write (see \cite{Varadarajan:2004yz}) $X_{ij}=X_{i}\cap X_{j}$,
$X_{ijk}=X_{i}\cap X_{j}\cap X_{k}$, and so on. We now introduce
isomorphisms of sheaves which represent the ``coordinate changes''
on our super-manifold. They allow us to glue the single pieces to
get the final supermanifold. Let
\[ f_{ij}:\left(X_{ji}, \mathcal{O}_{j}|_{X_{ji}} \right)\longrightarrow \left(X_{ij}, 
\mathcal{O}_{i}|_{X_{ij}} \right) \]
be an isomorphisms of sheaves with
\[ f_{ij}^{\sim}=Id.\]
This means that these maps represent differentiable coordinate
changes on the underlying manifold.

To say that we glue the ringed spaces $(X_{i},\mathcal{O}_{i})$ through
the $f_{ij}$ means that we are constructing a sheaf of rings $\mathcal{O}$
on $X$ and for each $i$ a sheaf isomorphism
\[ f_{i}:(X_{i},\mathcal{O}|_{X_{i}})\longrightarrow (X_{i},\mathcal{O}_{i}), \]
\[ f_{i}^{\sim}=Id_{X_{i}} \]
such that
\[f_{ij}=f_{i}f_{j}^{-1},\]
for all $i$ and $j$.

The following usual cocycle conditions are necessary and sufficient for
the existence of the sheaf $\mathcal{O}$:
\begin{description}
\item [{i.}] $f_{ii}=Id$ on $\mathcal{O}_{i}$;
\item [{ii.}] $f_{ij}f_{ji}=Id$ on $\mathcal{O}_{i}|_{X_{i}}$;
\item [{iii.}] $f_{ij}f_{jk}f_{ki}=Id$ on $\mathcal{O}_{i}|_{X_{ijk}}$.
\end{description}

\section{Projective superspaces} \label{proj}
Due to their importance in physical applications 
we now give a detailed description of projective superspaces. 
One can work either on $\mathbb{R}$ or on $\mathbb{C}$, but we choose
to stay on $\mathbb{C}$. Let $X$ be the complex projective space
of dimension $n$. The super-projective space will be called $Y$.
The homogeneous coordinates are $\left\{ z_{i}\right\} $. Let us
consider the underlying topological space as $X$, and let us construct
the sheaf of super-commutative rings on it. For any open subset $V\subseteq X$
we denote by $V^{\prime}$ its preimage in $\mathbb{C}^{n+1}\setminus\left\{ 0\right\} $.
Then, let us define $A\left(V^{\prime}\right)=H\left(V^{\prime}\right)\left[\theta^{1},\theta^{2},...,\theta^{q}\right]$,
where $H\left(V^{\prime}\right)$ is the algebra of holomorphic functions
on $V^{\prime}$ and $\left\{ \theta^{1}, \theta^{2}, ...,\theta^{q}\right\} $
are the odd generators of a Grassmann algebra. $\mathbb{C}^{\ast}$
acts on this super-algebra by: 
\begin{equation}
 t:{\sum_{I}}f_{I}\left(z\right)\theta^{I}\longrightarrow{\sum_{I}}t^{-|I|}f_{I}\left(t^{-1}z\right)\theta^{I}. \label{nonzero}\end{equation}
The super-projective space has a ring over $V$ given by:
\[ \mathcal{O}_{Y}\left(V\right)=A\left(V^{\prime}\right)^{\mathbb{C}^{\ast}} \]
which is the subalgebra of elements invariant by this action. This
is the formal definition of a projective superspace (see for example \cite{Varadarajan:note}), however we would like to construct the same space more explicitly from gluing different 
superdomains as in sec. \ref{smfld}.

Let $X_{i}$ be the open set where the coordinate $z_{i}$ does not
vanish. Then the super-commutative ring $\mathcal{O}_{Y}\left(X_{i}\right)$
is generated by elements of the type 
\[ f_{0}\left(\frac{z_{0}}{z_{i}}, \dots, \frac{z_{i-1}}{z_{i}}, \frac{z_{i+1}}{z_{i}}, \dots,\frac{z_{n}}{z_{i}}\right)\,,\quad f_{r}\left(\frac{z_{0}}{z_{i}},...,\frac{z_{i-1}}{z_{i}},\frac{z_{i+1}}{z_{i}},...,\frac{z_{n}}{z_{i}}\right)\frac{\theta^{r}}{z_{i}}\,,\quad r=1,\dots,q\,.\]

In fact, to be invariant with respect to the action of $\mathbb{C}^{\ast}$,
the functions $f_{I}$ in equation (\ref{nonzero}) must be homogeneous
of degree $-|I|$. Then, it is obvious that the only coordinate we
can divide by, on $X_{i}$, is $z_{i}$: all functions $f_{I}$ are
of degree $-|I|$ and holomorphic on $X_{i}$. If we put, on $X_{i}$,
for $l\neq i$, $\Xi_{l}^{(i)}=\frac{z_{l}}{z_{i}}$
and $\Theta^{(i)}_{r}= \frac{\theta^{r}}{z_{i}}$, then
$\mathcal{O}_{Y}\left(X_{i}\right)$ is generated, as a super-commutative
ring, by the objects of the form
\[F_{0}^{(i)}\left(\Xi_{0}^{(i)}, \Xi_{1}^{(i)},...,\Xi_{i-1}^{(i)}, \Xi_{i+1}^{(i)},...,\Xi_{n}^{(i)}\right), \quad F_{a}^{(i)}\left(\Xi_{0}^{(i)}, \Xi_{1}^{(i)},...,\Xi_{i-1}^{(i)}, \Xi_{i+1}^{(i)},...,\Xi_{n}^{(i)}\right)\Theta^{(i)}_{a}, \]
where $F_0^{(i)}$ and the $F_{a}^{(i)}$'s are analytic functions on $\mathbb{C}^{n}$. In order to avoid confusion we have put the index $i$ in parenthesis: it just denotes the fact that we are defining objects over the local chart $X_i$. In the following, for convenience in the notation,
we also adopt the convention that $\Xi^{(i)}_{i} = 1$ for all $i$.

To explain the ``coordinate change'' morphisms let us recall what happens in the ordinary 
complex projective spaces.

If we consider $\mathbb{P}^{n}(\mathbb{C})$ with the ordinary complex analytic
structure, then, over the affine open set $X_{i}$ where $z_{i}\neq0$,
we can define the affine coordinates $ w_{a}^{(i)}=\frac{z_{a}}{z_{i}}$,
$a \neq i$. The sheaf of rings over $X_{i}$ is $H(X_{i})$, the ring of analytic functions 
over $X_{i}$. Every element $f$ of $H(X_{i})$ can also be expressed as a function in 
homogeneous coordinates $F(z_{0},z_{1},...,z_{n})$. Two functions, $F^{(i)}$ on $X_i$ and
$F^{(j)}$ on $X_j$, represent ``the same function'' on the intersection $X_i \cap X_j$ if, 
when expressed in homogeneous coordinates, they give the same function $F$. 
The isomorphism between 
$(X_{i}\cap X_{j}, H(X_{i})|_{X_{j}})$ and $(X_{j}\cap X_{i}, H(X_{j})|_{X_{i}})$
sends $F^{(i)}$ to $F^{(j)}$, i.e. expresses $F^{(i)}$ with respect to the affine coordinates 
$w_{a}^{(j)}=\frac{z_{a}}{z_{j}}$. The total manifold is obtained by gluing these domains 
$X_{i}$ as in the previous section.

We now return to considering the super-projective spaces. We have
the two sheaves $\mathcal{O}_{Y}(X_{i})|_{X_{j}}$ and $\mathcal{O}_{Y}(X_{j})|_{X_{i}}$.
In the same way as before, we have the morphisms given by the ``coordinate
changes''. So, on $X_{i}\cap X_{j}$,
the isomorphism simply affirms the equivalence between the objects
of the super-commutative ring expressed either by the first system
of affine coordinates, or by the second one. So for instance we have that 
$\Xi_{l}^{(j)}=\frac{z_{l}}{z_{j}}$
and $\Theta^{(j)}_{r}=\frac{\theta^{r}}{z_{j}}$ can be also expressed as
\[ \Xi_{l}^{(j)}= \frac{\Xi_{l}^{(i)}}{\Xi_{j}^{(i)}}, \quad \Theta^{(j)}_{r} = \frac{\Theta_{r}^{(i)}}{\Xi_{j}^{(i)}}. \]
Which, in the language used in the previous section, means that the morphism $\psi_{ji}$ gluing $(X_{i}\cap X_{j}, \mathcal{O}_{Y}(X_{i})|_{X_{j}})$ and
$(X_{j} \cap X_{i}, \mathcal{O}_{Y}(X_{j})|_{X_{i}})$ is such that $\psi_{ji}^{\sim}$ is the usual change of coordinates map on projective space and 
 \[ \psi_{ji}^{\ast} (\Xi_{l}^{(j)}) = \frac{\Xi_{l}^{(i)}}{\Xi_{j}^{(i)}}, \quad \psi_{ji}^{\ast} (\Theta^{(j)}_{r}) = \frac{\Theta_{r}^{(i)}}{\Xi_{j}^{(i)}} \]

The super-manifold is obtained by observing that the coordinate changes
satisfy the cocycle conditions of the previous section.

\section{The functor of points}
We now wish to explain how the physicists' interpretation of the $z_{i}$'s as 
``even coordinates'' and the $\theta_{j}$'s as ``odd coordinates'' can be obtained from the ``super-ringed space'' interpretation of supermanifolds through the concept of ``functor of 
points''. The key to understanding this is Theorem~\ref{morphisms}.

Given two supermanifolds $X$ and $S$, the $S$-points of $X$ (or
the points of $X$ parametrized by $S$) are given by the set \[
X(S)={\rm Hom}(S,X)=\{{\rm set\,\, of\,\, morphisms\,\, S\rightarrow X\}\,.}\]
 $X$ is the supermanifold we want to describe and $S$ is the model
on which we base the description of $X$. Changing $S$ modifies the
description of $X$. The functor which associates $S$ to $X(S)$
is a functor between the category of supermanifolds and the category
of sets (which are the ``points'' of the supermanifolds). See also \cite{Fioresi:2006gx} for 
more details.

Let us interpret this in the case when $X=V^{r|s}$ and $S=U^{p|q}$.
According to Theorem~\ref{morphisms}, a morphism $\psi\in{\rm Hom}(U^{p|q},V^{r|s})$
is uniquely determined by a choice of $r$ even sections and $s$
odd sections of $C^{\infty\, p|q}(U)$, i.e. morphisms are in one
to one correspondence with $(r+s)$-tuples $(f_{1},\ldots,f_{r},g_{1},\ldots,g_{s})$,
where $f_{j}$'s are even and $g_{j}$'s are odd in the algebra $C^{\infty\, p|q}(U)$.
If we denote by $\Gamma_{q}^{0}(U)$ and $\Gamma_{q}^{1}(U)$ respectively
the set of even and odd sections of $C^{\infty\, p|q}(U)$, then the
above fact is expressed as \begin{equation}
{\rm Hom}(U^{p|q},V^{r|s})=(\Gamma_{q}^{0}(U))^{r}\times(\Gamma_{q}^{1}(U))^{s}.\end{equation}
 The sub-index $q$ denotes the {}``number of odd generators'' of
the algebra we are considering.

In particular, if $S=\mathbb{R}^{0|q}$, then 
\begin{equation}
{\rm Hom}(\mathbb{R}^{0|q},V^{r|s})=(\Gamma_{q}^{0})^{r}\times(\Gamma_{q}^{1})^{s}\end{equation}
 where $(\Gamma_{q}^{0})$ and $(\Gamma_{q}^{1})$ represent the even
and the odd component of a Grassmann algebra with $q$ generators,
respectively.

One could say that the {}``super-ringed space'' structure of $X$
encodes the information of how the even and odd coordinates $(z,\theta)$
glue together, but independently of the number of generators of the
underlying super-algebra. The number of generators ($q$ in the above
case) can be fixed by taking a supermanifold $S$ and constructing
${\rm Hom}(S,X)$. We will see some examples shortly.


\subsection{Coordinates of Superprojective Spaces.}

We are going to consider the superprojective space \begin{equation}
\mathbb{P}^{p|q}=\left(\mathbb{P}^{p},\mathcal{O}_{\mathbb{P}^{p}}\right)\end{equation}
 which is defined as in section \ref{proj} as a ringed space and
$\dim(\mathcal{O}_{\mathbb{P}^{p}})=q$. We want to describe
the set of $\mathbb{C}^{0|N}-$ points of this space. The space $\mathbb{C}^{0|N}$
can be viewed as the super-commutative ring ${\cal O}_{\mathbb{C}^{0}}$
over the single point (denoted by $P$) of the corresponding topological space 
$\mathbb{C}^{0}$ and can be identified precisely with the Grassmann algebra with $N$ generators 
that we denote by $\Gamma_{N}$.

Let's consider the open subsets $\left\{ X_{i}\right\} $, $i=0,1,2,...,p$,
of $\mathbb{P}^{p}$ where $z_{i}\neq0$, with the corresponding super-commutative
ring $\mathcal{O}_{X_{i}}$. A morphism between $\mathbb{C}^{0|N}$
and $(X_{i},\mathcal{O}_{X_{i}})$ is completely defined by the pull-back
for each generator of the ring ${\cal O}_{X_{i}}$ 
\begin{equation}
\tau_{(i)}\in{\rm Hom}(\mathbb{C}^{0|N},(X_{i},\mathcal{O}_{X_{i}}))\,,\quad\quad \tau_{(i)}^{*}:\mathcal{O}_{X_{i}}\rightarrow{\cal O}_{\mathbb{C}^{0}}\end{equation}
where ${\cal O}_{\mathbb{C}^{0}}=\mathbb{C}[\theta_{1},...,\theta_{N}]=\Gamma_{N}$.
To clarify this point, we take the generators of $\mathcal{O}_{X_{i}}$:
$\Theta^{(i)}_{j}\,,\quad j=1,\dots,q$ and the affine coordinates $\Xi^{(i)}_{j}$
on $X_{i}$ and we map into $\mathbb{C}[\theta_{1},...,\theta_{N}]$
as follows \begin{eqnarray}
\tau^{\ast}_{(i)}(\Xi^{(i)}_{j}) & = & f_j^{(i)} \quad j=1, \ldots, p \\
\tau^{\ast}_{(i)}(\Theta^{(i)}_{r}) & = & \eta_r^{(i)} \quad r=1,\dots,q\nonumber 
\end{eqnarray}
where the $f_j^{(i)}$ (resp. $\eta_r^{(i)}$) are even (resp. odd) elements of the Grassmann 
algebra $\Gamma_N$. It is clear that 
    $\tau_{(i)}^{\sim}(P) = ((f_1^{(i)})^{\sim}, \ldots, (f_p^{(i)})^{\sim})$. 
We therefore see that for every $i$, $\Hom(\mathbb{C}^{0|N},(X_{i},\mathcal{O}_{X_{i}}))$ can 
be identified with a copy of $(\Gamma^{0}_{N})^p \times (\Gamma^{1}_{N})^q$.

To obtain all the possible morphisms from $\mathbb{C}^{0|N}$ to $\mathbb{P}^{p|q}$,
we must take into account that the latter is built by ``gluing''
super-domains by means of the ``coordinate change isomorphisms'', this corresponds to gluing 
together all copies of $(\Gamma^{0}_{N})^p \times (\Gamma^{1}_{N})^q$ for all possible $i$'s.
Since a morphism in $\Hom (\mathbb{C}^{0|N},\mathbb{P}^{p|q})$
must be compatible with the restriction maps, it must commute with
the ``coordinate changes''. This means that, if $\tau_{(j)}^{\ast}$
is the pull-back of a morphism to $X_{j}$, and $\psi_{ij}^{\ast}:\mathcal{O}_{X_{j}}|_{X_{i}\cap X_{j}}\longrightarrow\mathcal{O}_{X_{i}}|_{X_{i}\cap{X_{j}}}$
is the isomorphism which represents ``coordinate changes'',
then
\[ \tau_{(j)}^{\ast}=\tau_{(i)}^{\ast}\circ \psi_{ij}^{\ast}. \]
This then induces a map between subsets of the $i$-th and $j$-th copy of 
$(\Gamma^{0}_{N})^p \times (\Gamma^{1}_{N})^q$ as follows
\[ (f_1^{(i)}, \ldots, f_p^{(i)}) \mapsto (f_j^{(i)})^{-1}(f_1^{(i)}, \ldots,1, 
                                                                     \ldots, f_p^{(i)}), \]
\[ (\eta_{1}^{(i)}, \ldots, \eta_q^{(i)}) \mapsto (f_j^{(i)})^{-1} 
                                       (\eta_{1}^{(i)}, \ldots, \eta_q^{(i)}). \]
By means of this map we glue the two copies together. Performing all these gluings gives 
a model for $\Hom (\mathbb{C}^{0|N},\mathbb{P}^{p|q})$, consisting of the $\mathbb{C}^{0|N}$-points of $\mathbb{P}^{p|q}$.

Another way to interpret this model is as follows.
We consider a set of ``homogeneous'' (even and odd) generators $z_{0},...,z_{p},\theta_{1},...,\theta_{q}$, where the $z_j$'s are in $\Gamma_N^{0}$ and at least one them is invertible and the 
$\theta_j$'s are in $\Gamma_N^{1}$. One obtains the local generators on each $X_{i}$ simply 
``dividing'' by $z_{i}$ (exactly like in the standard projective case, when
one looks for the ``affine coordinates'').
This way we see that we can identify 
\[ \Hom(\mathbb{C}^{0|N},\mathbb{P}^{p|q})=\frac{\left(\left(\Gamma_{N}^{0}\right)^{p+1}\setminus B^{p+1}\right)\times\left(\Gamma_{N}^{1}\right)^{q}}{\left(\Gamma_{N}^{0}\right)^{\ast}},\]
where $\left(\Gamma_{N}^{0}\right)^{\ast}$is the set of the even
invertible elements and $B=\left(\Gamma_{N}^{0}\right)\setminus\left(\Gamma_{N}^{0}\right)^{\ast}$. This model is exactly the generalization of the projective space as a supermanifold in the sense of Rogers, Bruzzo and others (see book \cite{Bruzzo} for  a complete discussion). 


\section{Supergroups and Superdeterminants}

As another illustration of the meaning of the functor of points we consider the case
of supergroups. For simplicity we will just look at the cases of $GL(1|1)$, $SL(1|1)$ and 
finally we will give another construction of the superprojective space as the quotient space $SU(n|m)/U(n-1|m)$.

Let us now consider the simplest case of supergroup $GL(1|1)$. As a supermanifold, $GL(1|1)$ is isomorphic
to the super-domain $U^{2|2}=(U^2, \mathcal C^{\infty 2|2})$, where $U^2 = (\mathbb{C}^*)^2$. If $(z_1,z_2)$
are the coordinates on $U^2$ and $\theta_1, \theta_2$ are the generators of the Grassmann algebra, it is convenient
to use the notation in matrix form
\begin{equation}
\left( \begin{array}{cc}
z_{1} & \theta_{1} \\
\theta_{2} & z_{2}
\end{array}\right).
\end{equation}

\medskip
We can define the ``product'' on $GL(1|1)$ as a morphism $$\psi \in \Hom(GL(1|1) \times GL(1|1), GL(1|1))$$ such that
$$
\begin{array}{ccc}
\psi^{\sim}:GL(1|1)_{0} \times GL(1|1)_0  &\longrightarrow &
GL(1|1)_0\\
 (z_{1}, z_{2}) \times (z_3, z_4) &\longmapsto & (z_1 z_3, z_2 z_4),
\end{array}
$$
$$
\begin{array}{ccc}
\psi^{*}: \mathcal{C}^{\infty}(U^2)\, [\vartheta_1,\vartheta_2] &\longrightarrow &
\mathcal{C}^{\infty}(U^2 \times U^2)\, [\theta_1,\theta_2,\theta_3,\theta_4]
  \\
\ \\
\left(
\begin{array}{cc}
w_{1} & \vartheta_{1} \\
\vartheta_{2} & w_{2}
\end{array}\right)
&\longmapsto&
\left(
\begin{array}{cc}
z_{1} z_{3} + \theta_{1} \theta_{4} & \theta_{1} z_{4} + z_{1} \theta_{3} \\
 \theta_{2} z_{3} + z_{2} \theta_{4} & z_{2} z_{4} + \theta_{2} \theta_{3}
\end{array}\right),
\end{array}
$$
where the action of the pull-back morphism $\psi^*$ has been specified only for the generators
of the algebra (see Theorem~\ref{morphisms}).

We now apply the functor of points to recover the usual interpretation of $GL(1|1)$ as the
set of ``invertible supermatrices''. Take as model space $S = \mathbb C^{0|q}$, then $\Hom(S, \GL(1|1))$
can be identified with the set of matrices
\begin{equation}\label{gl}
g= \left(
\begin{array}{cc}
\psi^*z_{1} & \psi^*\theta_{1} \\
\psi^* \theta_{2} & \psi^* z_{2}
\end{array}\right),
\end{equation}
where $\psi^*z_{i}$ are even elements of the Grassmann algebra $\Gamma_q$, whose value is
different from zero, and $\psi^*\theta_{i}$ are odd elements. To simplify notation we denote
$\psi^* z_i$ (resp. $\psi^* \theta_i$) by $z_i$ (resp. $\theta_i$).

The above ``product''  becomes the usual multiplication of super-matrices as follows. A morphism form
$S$ to $\GL(1|1) \times \GL(1|1)$ is given by a pair of matrices, $g_1$ and $g_2$, as above. 
Composition with the product morphism gives a morphism from $S$ to $\GL(1|1)$, represented by 
a matrix $g_3$, which can be seen to be given by the usual multiplication of matrices:
\begin{equation}\label{glB}
g_1 =\left(
\begin{array}{cc}
z_{1} & \theta_{1} \\
\theta_{2} & z_{2}
\end{array}\right) \,,
\hspace{1cm}
g_2 =
\left(
\begin{array}{cc}
z_{3} & \theta_{3} \\
\theta_{4} & z_{4}
\end{array}\right)
\end{equation}
$$
g_3 = g_1 g_2 =
\left(
\begin{array}{cc}
z_{1} z_{3} + \theta_{1} \theta_{4} & \theta_{1} z_{4} + z_{1} \theta_{3} \\
 \theta_{2} z_{3} + z_{2} \theta_{4} & z_{2} z_{4} + \theta_{2} \theta_{3}
\end{array}\right).
$$

Recall the classical formula for the superderminant (or Berezinian) of a super-matrix in 
$\GL(1|1)$:
\begin{equation}\label{ber}
sdet(g) = Ber(g) = \frac{z_{1}}{z_{2}}\left( 1 + \frac{\theta_{1} \theta_{2}}{ z_{1} z_{2}}\right)
\end{equation}
which is well defined if $z_{2}\neq 0$.
The Berezinian can also be understood from the sheaf point of view, as a morphism $Ber$ from
$GL(1|1)$ to $\mathbb{C}^{1|0}$:
$$
\begin{array}{ccc}
Ber^{\sim}:GL(1|1)_{0}  &\longrightarrow & \mathbb{C}
\\
(z_{1}, z_{2}) &\longmapsto &{z_{1}}/{z_{2}}
\end{array}\,, ~~~~~~
\begin{array}{ccc}
Ber^{*}: \mathcal C^{\infty 1|0} &\longrightarrow &
\mathcal C^{\infty 2|2} \\
w &\longmapsto&
\frac{z_{1}}{z_{2}}\left(1+\frac{\theta_{1} \theta_{2}}{ z_{1} z_{2}}\right) \,.
\end{array}
$$

Next, we consider a subset of supermatrices $GL(1|1)$ with the property that
``the superdeterminant is 1''. They are denoted by $SL(1|1)$.
We want to describe this space using the sheaf theoretic interpretation of supermanifolds, by 
restricting the base manifold and considering an appropriate quotient sheaf.
We need to give a meaningful interpretation of the relation
\begin{equation}\label{berC}
\frac{z_{1}}{z_{2}}\left( 1 + \frac{\theta_{1} \theta_{2}}{ z_{1} z_{2}} \right) = 1\,.
\end{equation}
We do it as follows. Let $\mathcal{J}$ be the ideal in $\mathcal{C}^{\infty \, 2|2}$ generated 
by
\begin{equation}\label{berC-1}
\frac{z_{1}}{z_{2}}\left( 1 + \frac{\theta_{1} \theta_{2}}{ z_{1} z_{2}} \right) - 1\,.
\end{equation}
The base manifold of $SL(1|1)$ is the support of this ideal, i.e. the subset 
$X \subset \mathbb{C}^* \times \mathbb{C}^*$ of points around which no element of $\mathcal{J}$ 
is invertible. Clearly $X$ is the diagonal
in $\mathbb{C}^* \times \mathbb{C}^*$, i.e. the set where $z_1 = z_2$. 
The sheaf $\mathcal{O}_X$ is the restriction to $X$ of the quotient sheaf
\begin{equation}\label{slsheaf}
\mathcal{C}^{\infty \, 2|2}/\mathcal{J}.
\end{equation}
It remains to show that this ringed manifold $SL(1|1)$ is really a
supermanifold, i.e. it is obtained by pasting super-domains. In fact observe that the relation 
(\ref{berC-1}) tells us that over an open set $V \subset X$, $z_1 = z_2 - \frac{\theta_1 \,
\theta_2}{z_2}$. Therefore the ring over $V$ is
$\mathcal{C}^{\infty} (z_2 - \frac{\theta_1 \,
\theta_2}{z_2}, z_2)[\theta_1,\theta_2]$ which can be seen to be isomorphic 
to a super-ring of the type $\mathcal{C}^{\infty}(z)[\Psi_1,\Psi_2]$, so locally $SL(1|1)$ is 
isomorphic to a superdomain. What we have done here is
to show explicitly that $SL(1|1)$ is a sub-supermanifold of dimension $1|2$ of $\GL(1|1)$ in the sense of \cite{Varadarajan:note}. 

To conclude the description of $X=SL(1|1)$, we present its
interpretation by means of the functor of points, using the model
space $S=\mathbb{C}^{0|N}$. Then, the morphisms in $\Hom(S,X)$ can
be viewed as the morphisms in $\Hom(S, GL(1|1))$ such that:
\begin{itemize}
\item[i.) ] the map between the underlying topological spaces has
image contained in the diagonal of $\mathbb{C}^* \times
\mathbb{C}^*$;
\item[ii.) ] the pull-back map descends to the quotient, i.e.
$\psi^*(j)=0$ for any $j \in \mathcal{J}$.
\end{itemize}
Then, the set $\Hom(S,X)$ can be viewed as the matrices of the form
\begin{equation}\label{sl}
g= \left(
\begin{array}{cc}
\psi^*z_{1} & \psi^*\theta_{1} \\
\psi^* \theta_{2} & \psi^* z_{2}
\end{array}\right),
\end{equation}
with the conditions that: 1) the value of $\psi^*z_{1}$ is equal to
the value of $\psi^*z_{2}$ (by (i)), and 2) the super-determinant of
the matrix in (\ref{sl}) is $1$ (by (ii)).

We start describing $GL(1|1,\mathbb{C})$ in a different way, by passing to
real super-groups. We use the following idea: think about $\mathbb{C}^{n}$, with a
complex basis $\left\{ v_{1};v_{2};...;v_{n}\right\} $. Then, $\mathbb{C}^{n}$ 
can be viewed as a real $2n$-dimensional space, with a real basis
\begin{center}
$\left\{ v_{1};iv_{1};v_{2};iv_{2};...;v_{n};iv_{n}\right\}\,.$
\end{center}

So, let's take $V^{2}=\left\{ (a;b)\in \mathbb{R}^{2}:a^{2}+b^{2}\neq
0\right\} $. We take, as a base manifold, $\left( V^{2}\right) ^{2}$, and the
ring over this total space is $\mathcal{C}_{\mathbb{R}}^{\infty }\left(
z_{1};z_{1}^{\prime };z_{2};z_{2}^{\prime }\right) \left[ \theta ;\theta
^{\prime };\psi ;\psi ^{\prime }\right] $. Note that $z_{1}^{\prime }$ and 
$z_{2}^{\prime }$ (as even generators), and $\theta ^{\prime }$ and 
$\psi^{\prime }$ (as odd generators) play the role of the vectors (written above)
$iv_{j}$. So, the primed generators are independent from the real point of
view; sums like $xz_1+yz^{\prime}_1$ (or $\theta + \theta^{\prime}$)
represent complex even (respectively, odd) elements, since they put together
the real and the imaginary part.

We can view the generators in the following matrix form, which helps us in
writing the product morphism.
\begin{equation}  \label{u11}
\left(
\begin{array}{cc}
z_{1}+z_{1}^{\prime } & \theta +\theta ^{\prime } \\
\psi +\psi ^{\prime } & z_{2}+z_{2}^{\prime }
\end{array}
\right).
\end{equation}
\bigskip

The product morphism is defined by:
$m:$ $\left( V^{2}\right) ^{2}\times $ $\left( V^{2}\right)
^{2}\longrightarrow $ $\left( V^{2}\right) ^{2}\,,$
and one can describe exactly the pull-back morphism by performing
explicitly the matrix multiplication:
$$\left(
\begin{array}{cc}
z_{1}^{1}+z_{1}^{\prime 1} & \theta ^{1}+\theta ^{\prime 1} \\
\psi ^{1}+\psi ^{\prime 1} & z_{2}^{1}+z_{2}^{\prime 1}
\end{array}
\right) \cdot \left(
\begin{array}{cc}
z_{1}^{2}+z_{1}^{\prime 2} & \theta ^{2}+\theta ^{\prime 2} \\
\psi ^{2}+\psi ^{\prime 2} & z_{2}^{2}+z_{2}^{\prime 2}
\end{array}
\right) $$
Then, we try to define a ''complex conjugation''.
It is a morphism, such that its pull-back map takes the form:
$$\rho ^{\ast }:\mathcal{C}_{\mathbb{R}}^{\infty }\left( z_{1};z_{1}^{\prime
};z_{2};z_{2}^{\prime }\right) \left[ \theta ;\theta ^{\prime };\psi ;\psi
^{\prime }\right] \longrightarrow \mathcal{C}_{\mathbb{R}}^{\infty }\left(
z_{1};z_{1}^{\prime };z_{2};z_{2}^{\prime }\right) \left[ \theta ;\theta
^{\prime };\psi ;\psi ^{\prime }\right]\,,$$
and it is completely determined once one knows the behavior of the
generators. In fact, it sends $z_{1};z_{2};\theta ;\psi $ to themselves,
while the primed elements $z_{1}^{\prime };z_{2}^{\prime };\theta ^{\prime
};\psi ^{\prime }$ undergo a change of sign, going respectively to 
$-z_{1}^{\prime };-z_{2}^{\prime };-\theta ^{\prime };-\psi ^{\prime }$.

Next, we define the ''Hermitian transpose'', which will be denoted
by the symbol $\mathcal{H}$.
Its pull-back map sends $z_{1};z_{2}$ to themselves, $z_{1}^{\prime
};z_{2}^{\prime }$ to $-z_{1}^{\prime };-z_{2}^{\prime }$, respectively; 
$\theta ;\theta^{\prime}$ to $\psi^{\prime};\psi $, respectively, and 
$\psi^{\prime};-\psi$ to $\theta^{\prime};-\theta$, respectively. Note that
the exchange of the $\theta$ with the $\psi$ is due to the transposition
operation, while the exchange of a non-primed generator with a primed one is
due to a multiplication by $-i$. We will see this better when we pass to
Grassmann algebras by the functor of points.
We have again to require that $\mathcal{H}$ is a morphism. Note that it is
not true that $\left( \mathcal{H}^{\ast }\right) ^{2}={\rm Id}\,.$

Now, we can describe the functor of points with respect to a particular
model space. We choose, as a model space, $\mathbb{C}^{0|q}$.
So, $\mathbb{C}^{0|q}=\left( \left\{ pt.\right\} ;\mathbb{C}\left[ \xi
_{1};\xi _{2};...;\xi _{q}\right] \right) $. The conjugation is a map,
defined, in a standard way (since $\mathbb{C}^{0|q}$ is simply a Grassmann
algebra)
$\sigma :\mathbb{C}^{0|q}\longrightarrow \mathbb{C}^{0|q}$, with pull-back
defined by:
$\sigma ^{\ast }\left( \left( x+iy\right) \xi _{i}\right) =\left(
x-iy\right) \xi _{i}$.

When we construct the morphisms $\overline{Hom}\left( \mathbb{C}%
^{0|q};GL\left( 1|1\right) \right) $, we only take the ones which are $%
\mathbb{R}$-linear and compatible with conjugations, which means that $%
\sigma ^{\ast }\circ \varphi^{\ast }=\varphi^{\ast }\circ \rho^{\ast }$.
This means for example, that, if we choose an odd element of $\mathbb{C}%
^{0|q}$ as the pull-back of $\theta $, the pull-back of $\theta ^{\prime }$
is consequently $i$ times the pull-back of $\theta $. (In general this is not true, but 
we have assumed here that there is a complex structure $J$ on the anticommuting coordinates and 
we consider only those morphisms which commute with $J$.) 

In this way, we can describe the $\mathbb{C}^{0|q}$-points of $GL\left(
1|1\right) $ as the set of matrices of the form
\begin{center}
$\left(
\begin{array}{cc}
a & \alpha  \\
\beta  & b
\end{array}
\right) $,
\end{center}
where $a;b$ are even elements of a complex Grassmann algebra with $q$
generators, $\alpha $;$\beta $ are odd elements of the same algebra, and the
Berezinian is invertible.

Now, we can pass to construct the super-group $U(1|1).$ Note that it is a
real super-manifold.
We have to perform the quotient of $\mathcal{C}_{\mathbb{R}}^{\infty }\left(
z_{1};z_{1}^{\prime };z_{2};z_{2}^{\prime }\right) \left[ \theta ;\theta
^{\prime };\psi ;\psi ^{\prime }\right] $, by the four relations obtained
from:
\begin{equation}
A\cdot \mathcal{H}^{\ast }\left( A\right) =\mathbb{I},
\end{equation}
where $A$ is a sheaf element written in the matrix form (\ref{u11}). By
explicitly writing the elements on the left, we get four generators of an
ideal $I$. Taking the quotient
$${\mathcal{C}_{\mathbb{R}}^{\infty }\left(
z_1;z^{\prime}_1;z_2;z^{\prime}_2\right) \left[ \theta ;\theta^{\prime};\psi
;\psi^{\prime}\right] }/{I}\,,$$ we finally get the sheaf corresponding to 
$U(1|1)$. The base manifold is the support of this sheaf. 
Applying the functor of points interpretation and using $\mathbb{C}^{0|q}$
as a model, it is easy to see that the elements of $U(1|1)$ correspond, as a
set, to the matrices of the form
\begin{equation}
B=\left(
\begin{array}{cc}
a & \alpha  \\
\beta  & b
\end{array}
\right)\,.
\end{equation}
They preserve the "scalar" product  $\langle (z, \theta), (z, \theta) \rangle = 
z \bar z + i \theta \bar\theta $. Note that the bar is the conjugation in $\mathbb{C}^{0|q}$, i.e.
$\overline{a}=\sigma ^{\ast }\left( a\right) $.
It follows that 
$B^{\dag }\cdot B=\mathbb{I}$, where $^{\dagger }$ represents the
usual ''adjoint'' of super-matrices. It represents the correspondence
between a matrix and its Hermitian transpose from the $\mathbb{C}^{0|q}$
point of view, since we can see that for every element $\varphi \in
\overline{Hom}\left( \mathbb{C}^{0|q};GL\left( 1|1\right) \right) $, $%
\varphi ^{\ast }\circ $ $\mathcal{H}^{\ast }$ is related to $\varphi ^{\ast }
$ by exactly performing the $^{\dagger }$ operation.
More precisely,
\begin{equation}
B^{\dagger }=\left(
\begin{array}{cc}
\overline{a} & -i\overline{\beta } \\
-i\overline{\alpha } & \overline{b}
\end{array}
\right)\,.
\end{equation}

For the sake of completeness, we write explicitly the $\mathbb{C}^{0|q}$-points 
of $U(1|1)$. They are in bijective correspondence with the matrices
of the form
\begin{equation}
U=\left(
\begin{array}{cc}
1-\frac{i}{2}\gamma \overline{\gamma } & -ie^{i\psi }\overline{\gamma } \\
\gamma  & e^{i\psi }\left( 1+\frac{i}{2}\gamma \overline{\gamma }\right)
\end{array}
\right) ,
\end{equation}
where $\psi $ is a real phase and $\gamma $ is a generic odd element of the
Grassmann algebra. 

A similar construction applies to $U(n|m)$ supergroups.
To get the supergroups $SU(n|m)$ we have to quotient with respect to the Berezinian equal to one.  The body part of $SU(n|m)$ is 
$U(1) \times SU(n) \times SU(m)$. The odd part belongs to 
the fundamental representation of $SU(n) \times U(m)$.


\subsection{Superprojective spaces as Supercosets}

Here we show that, using the functor-of-points framework, the superprojective 
spaces can be described in three different and equivalent ways. We first remind the 
reader the three methods to define the classical projective space and then we 
extend it to superprojective ones. 

\def\PN{$\mathbb{P}^n$}

In the classical case, 
let $z_{i}, i=1, \dots, n+1$ be the coordinates on $\mathbb{C}^{n+1}- \{ 0 \}$
and define the projective space $\mathbb{P}^n$ by the quotient
\begin{equation}
\label{newA}
(z_{1},\dots, z_{n+1}) \sim \lambda (z_{1},\dots, z_{n+1})\,, ~~~~
\lambda \in \mathbb{C}^{*}\,. 
\end{equation}
This is the standard definition of  \PN. Alternatively, one can fix 
the modulus of $\lambda$ by setting 
\begin{equation}
\label{newB}
\sum_{i=1}^{n+1} |z_{i}|^{2} = r > 0\,, ~~~~~
z_{i} \sim  e^{i \phi}z_{i}\,, ~~~\forall i\,, ~~~~
\phi \in {\mathbb R}\,.
\end{equation}
up to the phase $\phi$. 
The first equation fixes the modulus $|\lambda|^{2} =1$ and
the second equation removes its phase. Let us choose $r =1$, this
implies that the vector $z_{i}$ has modulus equal to one.

The $SU(n+1)$ symmetry of (\ref{newB}) is used to
bring the vector $z_{i}$ in the form
$(1,0,\dots,0)$ which has modulus equal to one. 
This vector has a stability group
which is $U(n)$. A stability group is the subgroup of transformations
which leaves $(1,0, \dots,0)$ invariant. Therefore, we can
define the projective space as the coset
\begin{equation}
\label{newC}
SU(n+1)/U(n)\,.
\end{equation}
The three ways to define a \PN\ are easily seen to be equivalent. 

\def\PNM{$\mathbb{P}^{n|m}$ }

Let us consider now the superprojective spaces 
$\mathbb{P}^{n|m}$.
The definition (\ref{newA}) can be repeated as follows
\begin{equation}
\label{suA}
(z_{1},\dots, z_{n+1}, \theta_{1}, \dots, \theta_{m})
\sim \lambda (z_{1},\dots, z_{n+1}, \theta_{1}, \dots, \theta_{m})\,, ~~~~
\end{equation}
with $\lambda$ not belonging to $\mathbb{C}^{*}$, but to the space of
even quantities which are invertible (see sec. 4.1).
Again, we can use the alternative definition of \PNM 
(\cite{Schwarz:1995ak,Konechny:1997hr,Witten:2003nn,Varadarajan:note})
\begin{equation}
\label{suB}
\sum_{i=1}^{n+1} |z_{i}|^{2}  + i \, \sum_{A=1}^{m} \bar \theta_{A}
\theta_{A} = r > 0\,, ~~~~~
z_{i} \sim e^{i \phi} z_{i}\,, ~~~\theta_{A} \sim e^{i \phi} \theta_{A}\,,~~~~~
\forall i,A\,, ~~~~
\phi \in \mathbb{C}\,.
\end{equation}

The equation (\ref{suB}) is the correct extension of (\ref{newB}),
but its interpretation needs some comment. As we have seen there are two ways to 
describe supermanifolds: {\it i}) using the sheaf description and {\it ii}) using the 
functor-of-points. According to the first framework, eq. (\ref{suB}) can be seen 
as an algebraic equation among the generators of the sheaves of the supermanifolds. 
Eq. (\ref{suB}) is consistent with the projection (\ref{suA}) and it defines an hypersurface 
in \PNM. According to the second description, one has to decompose the coordinates 
$z_i$ and $\theta_A$ on the basis of the generators of the superdomain and the coefficients 
need to satisfy a set of algebraic equations.

 Note that $SU(n+1|m)$ acts on \PNM as follows: 
 \begin{equation}\label{suACT}
\psi: SU(n+1|m) \times \mathbb{P}^{n|m} \longrightarrow \mathbb{P}^{n|m}
\end{equation}
with the pull-back defined by 
\begin{equation}
\psi^*(z_i) = \sum_{j=1}^{n+1} A_{ij} z_j + \sum_{A=1}^{m} \alpha_{i A} \theta_A\,, 
\hspace{2cm}
\psi^*(\theta_A) = \sum_{j=1}^{n+1} \beta_{A j} z_j + \sum_{B=1}^{m} B_{B A} \theta_I\,, 
\end{equation}
where $A_{ij}, B_{AB}$ are the even generators of $SU(n+1|m)$ and 
$\alpha_{i A}, \beta_{A j}$ are the odd ones. It is easy to see that the action 
is transitive like in the classical case. So, as in the classical case one  can define 
the supercoset $SU(n+1|m)/SU(n|m)$ which can be identified with the superprojective space $\mathbb{P}^{n|m}$. An analysis of the supercosets can be found in 
in the book \cite{manin} and recently it has been discussed in \cite{Fioresi:2006gx}.

Therefore applying the construction above and
starting from a vector $((1,\dots, 0), (0,\dots, 0))$ (where the first
set of components are the even coordinates and the second set the
odd ones) we end up with eq. (\ref{suB}). The odd part
of (\ref{suB}) is obtained by acting with the odd part of the
supergroup  on the unit vector. Notice that this is not
the only possibility, indeed we can start from an odd vector
 $((0,\dots, 0), (1,\dots, 0))$ which has the following norm
$$
|| ((0,\dots, 0), (1,\dots, 0)) ||^{2} =i\,  \theta_{1} \bar \theta_{1}\,.
$$
In this case, acting with the supergroup on it (and preserving
the subgroup $U(n|m)$), we end up with the new equation
\begin{equation}
\label{suD}
\sum_{i=1}^{n+1} |z_{i}|^{2}  +i\,  \sum_{A=1}^{m} \bar \theta_{A}
\theta_{A} = r\,,
\end{equation}
where $r$ is an even element of the algebra. For example,
starting from the vector $$((1,\dots, 0), (1,\dots, 0))$$ we have
$r = r_{0} +i\, \theta_{1} \bar \theta_{1}$ whose body
$r_{0}$ is positive.


\section{Balanced Supermanifolds}

In this section, we propose a possible extension of notion of {\it balanced manifold} 
(see \cite{dona}) to the supermanifolds. We found appropriate 
to report the present results since they call for a functor-of-point interpretation and 
for the definition of stable supermanifolds. 

\subsection{Donaldson's balanced superprojective spaces}

Let us consider the superprojective space $\mathbb{P}^{p|q}$ with 
standard coordinates $[z_{0}, \dots, z_{p}, \theta_{1}, \dots, \theta_{q}]$ and 
the matrix valued function on  $\mathbb{P}^{p|q}$ given by 
\begin{eqnarray}\label{nuovoA}
&&\hspace{3cm}
B_{ik} = \frac{z_{i} \bar z_{k}}{ \sum_{l=0}^{p} |z_{i}|^{2} + i \sum_{l=1}^{q} \theta_{I} \bar \theta_{I}}\,,  \nonumber \\
&&B_{i K} = \frac{z_{i} \bar \theta_{K}}{ \sum_{l=0}^{p} |z_{i}|^{2} + i 
\sum_{l=1}^{q} \theta_{I} \bar \theta_{I}}\,, ~~~~
B_{I k} = \frac{\theta_{I} \bar z_{k}}
{\sum_{l=0}^{p} |z_{i}|^{2} + i \sum_{l=1}^{q} \theta_{I} \bar \theta_{I}}\,,
 \\
&&
\hspace{3cm}
B_{I K} = \frac{i \theta_{I} \bar \theta_{K}}{\sum_{l=0}^{p} |z_{i}|^{2} + i \sum_{l=1}^{q} \theta_{I} \bar \theta_{I}}\,, \nonumber 
\end{eqnarray}
If we denote by $V$ a projective subsupervariety of $\mathbb{P}^{p|q}$, 
we define the  $(p+q +1) \times (p+q+1)$-matrix by the block matrix 
\begin{equation}\label{nuovoBB}
M(V)_{A B} =
\left(
\begin{array}{cc}
 \int_{V} B_{ik} d\mu_{V} &   \int_{V} B_{i K} d\mu_{V} \\
 \int_{V} B_{I k} d\mu_{V} & \int_{V} B_{I K} d\mu_{V}   
\end{array}
\right)
\end{equation}
where the indices $A, B$ run over $p+1+q$ values. Notice that 
$\overline{M(V)}_{i j} = M_{j i}$, $\overline{M(V)}_{i K} = M(V)_{I k}$ and $\overline{M(V)}_{I K} = 
- M_{K I}$. The measure 
$d\mu_{V}$ is defined as follows. For the superprojective space 
$\mathbb{P}^{p|q}$, the Fubini-Study form is given by 
\begin{equation}\label{nuovoC}
\Omega_{FS} = \frac{i}{2\pi} \partial\bar\partial \log\Big( 
\sum_{l=0}^{p} |z_{i}|^{2} + i \sum_{l=1}^{q} \theta_{I} \bar \theta_{I}
\Big)\,,
\end{equation}
where $\partial = dz^{i} \partial_{i} + d\theta_{I} \partial_{I}$ and 
$\bar\partial$ is its conjugate. The supertangent space and the cotangent space are defined in \cite{Varadarajan:note}. The expressions for $\partial$ and $\bar\partial$ are the natural extensions of usual geometry and acting on a superfunction they produce a 
superform of type $(1,1)$. The Fubini-Study form is real $\overline \Omega_{FS} = \Omega_{FS}$. 
Then, one can form 
\begin{equation}\label{nuovoD}
d\mu_{V} = \frac{\Omega^{p+q}_{FB}}{(p+q)!}\Big|_{V} = 
f_{V}(z, \bar z, \theta, \bar\theta) \, d^{p}z \wedge d^{p}\bar z\, \wedge d^{q}\theta \wedge d^{q}\bar\theta\,, 
\end{equation}
where $p+q$ is the sum of the bosonic and fermionic dimension of $V^{p|q}$ and 
$f_{\mu}(z, \bar z, \theta, \bar\theta)$ is a real superfield. 
 
Notice that the wedge product in $d\mu_{V}$ is a skew product 
for one-forms $dx^{i}$, while is a symmetric product for the $d\theta_{I}$'s. 
This super-$p+q$-form does no correspond to a integration 
measure for the supermanifold $\mathbb{P}^{p|q}$ since it is not 
clear how to integrate on the cotangent bundle generated by the commuting 
superforms $d\theta$ and $d\bar\theta$.
While the integration of functions on the supermanifold is clear 
since it is obtained by the Berezin integral \cite{Varadarajan:2004yz}, 
the integration on the superforms is obtained by using the method of the projection 
forms  
\begin{equation}\label{thom}
d \widetilde{\mu_{V}} = \frac{1}{p!}\Omega^p_{FB} \wedge {\cal U}_q
\end{equation}
where $ {\cal U}_m $ is the Thom class obtained by viewing the supermanifold 
 $\mathbb{P}^{p|q}$ as modeled on $\mathbb{P}^{p}$. For a 
 nice review see \cite{Cordes:1994fc}. The construction of 
 $ {\cal U}_m$ for  $\mathbb{P}^{p|q}$ will be given in \cite{mare}. 

Applying this rules, one immediately gets 
\begin{equation}\label{nuovoB}
M(V)_{i K} = M(V)_{I k} =0
\end{equation}
and there are only non-trivial blocks $M_{ik}$ and $M_{IK}$. Since the 
computation of ${\cal U}_{q}$ requires new ingredients, we consider in the following 
only super-Calabi-Yau spaces. For them we can use a different measure provided 
by the holomorphic form. 

Following Donaldson \cite{dona}, we define a balanced supermanifold if 
$M(V)$ is a multiple of the identity matrix. Notice that the identity matrix in 
the supermanifold has the block structure $\mathbb{I} = (\delta_{i \bar j}, \delta_{I \bar K})$. Hence, a supermanifold is balanced iff there exist two real numbers $\lambda$ and $\eta$ such that 
\begin{equation}\label{nuovoE}
\int_{V} B_{i \bar j} \, d\mu_{V} = \lambda \, \delta_{i \bar j}\,, ~~~~~~~
\int_{V} B_{i \bar J} \, d\mu_{V} = 0\,, ~~~~~~~
\int_{V} B_{I \bar J} \, d\mu_{V} = \eta \, \delta_{I \bar J}\,. ~~~~~~~
\end{equation}
If $p=q$, then we must have $\lambda=-\eta$. This is due to the presence 
of a additional $U(1)$ subgroup of the stability group $SU(p|q)$ (which 
is the group of isometries of the supermetric $\mathbb{I}$) and this reduces 
the supergroup to $PSU(p|p)$. 
Notice that the integration over the fermionic coordinates 
produces two terms: one is coming from the expansion of the denominator in 
$B_{i \bar j}$ or $B_{I \bar K}$ and the second is coming from the 
expansion of the measure $d\widetilde \mu_V$. 
The second source of interest in (\ref{nuovoE}) is the presence of the additional constraints (the second and the third relations)  
on the bosonic manifold.

In the case of super-Calabi-Yau space we use the integration  
measure obtained by the nowhere-vanishing holomorphic form $\Omega_{CY}$. 
This simplifies the construction and we give here the prescription how to integrate 
a given function $F$ (notice that in \cite{dona2}, in the case of CY's, the holomorphic form is used to define the measure in order to accelerate the convergence of integrals). We focus on the super-Calabi-Yau $\mathbb{P}^{1|2}$ whose 
holomorphic form is (see \cite{Witten:2003nn}) 
\begin{equation}\label{CYA}
\Omega_{CY} = z dz \epsilon_{ij} \frac{\partial}{\partial \theta^i} 
\frac{\partial}{\partial \theta^j}\,.
\end{equation}
Therefore, we have 
\begin{equation}\label{inte}
\int_{\mathbb{P}^{1|2}} 
\Omega_{CY}\wedge \bar\Omega_{CY} F(z,\bar z, \theta^i, \bar\theta^i) = 
\int_{\mathbb{P}^{1}}   |z|^2 dz\wedge d\bar z  \epsilon_{ij} 
\frac{\partial}{\partial \theta^i} 
\frac{\partial}{\partial \theta^j} \epsilon_{ij} 
\frac{\partial}{\partial \bar\theta^i} 
\frac{\partial}{\partial \bar\theta^j} 
F(z,\bar z, \theta^i, \bar\theta^i) |_{\theta=\bar\theta=0}\,.
\end{equation}
In the second line we have taken the four derivatives with respect to the 
fermionic coordinates $\theta_i$ and $\bar\theta_i$ and then set them to zero. 
It remains to perform the usual integration on the $\mathbb{P}^1$. 

Let us now consider a generic polarized supermanifold $(M,L)$ with 
$L$ an holomorphic super line bundle where the transition functions are elements of 
$GL(1|1)$. In addition, we require that the super line bundle 
has its first Chern class $c_{1}(L)$ represented by a K\"ahler form 
of the supermanifold $\omega$. For a positive $m$, we construct the tensor 
power series $L^{\otimes m}$ of the super line bundle and we denote by 
$H^{0}(M, L^{\otimes m})$ the space of holomorphic sections (as clarified above) 
of $L^{\otimes m}$. The 
holomorphic section could be odd or even. We use the 
extension of the Kodaira embedding theorem \cite{lebrun} asserting 
that for a sufficiently large $m$ the holomorphic sections define a 
projective embedding 
\begin{equation}\label{nuovoG}
i_{m}: M \rightarrow \mathbb{P}\Big(H^{0}(M,L^{\otimes m})\Big)
\end{equation}
A choice of holomorphic sections $(s_{0|0}, \dots, s_{p_{m}|q_{m}})$ in 
$H^{0}(M, L^{\otimes m})$ identifies 
$\mathbb{P}\Big(H^{0}(M,L^{\otimes m})\Big)$ with a superprojective 
$\mathbb{P}^{p_{m}|q_{m}}$ where the superdimensions $p_{m}$ and 
$q_{m}$ are due to the choice of even and odd sections. 

Let us consider an example. We consider $\mathbb{P}^{1|2}$ 
(which is a super-Calabi-Yau). 
Chosen $m$, $H^{0}(\mathbb{P}^{1|2}, L^{\otimes m})$ 
is spanned by 
\begin{equation}\label{nuovoH}
H^{0}(\mathbb{P}^{1|2}, L^{\otimes m}) = \Big\{ 
z^{a}_{0} z^{m -a}_{1}, z^{a}_{0} z^{m -a -1}_{1} \theta_I, 
z^{a}_{0} z^{m -a - 2}_{1} \theta_1 \theta_2
\Big\}
\end{equation}
$p_{m} = 2 m -1$ and $q_{m} = 2 m$. We can see that each space 
$H^{0}(\mathbb{P}^{1|2}, L^{\otimes m})$ is again a super-Calabi-Yau 
space. This can be verified easily using the formulas given in 
\cite{Grassi:2006cd} and it amounts to see that the number of anticommuting 
coordinates must exceed of one w.r.t. the commuting ones. 
In analogy with the bosonic case, we define the supermanifold $(M, L^{\otimes m})$ superbalanced if one can choose a basis in $\mathbb{P}\Big(H^{0}(M,L^{\otimes})\Big)$ such that the $V = \iota_{m}(M)$ is a superbalanced variety. 

On the space $\mathbb{P}\Big(H^{0}(M,L^{\otimes m})\Big)$ we can 
define the Kh\"aler form induced by the Fubini-Study form on 
$\mathbb{P}^{p_{m}|q_{m}}$, namely
\begin{equation}\label{nuovoI}
\omega_{m} = \frac{i}{2\pi} \partial\bar\partial \log 
\sum_{l=0}^{p_{m} + q_{m}} \Big|\frac{s_{l}(x)}{ \sigma(x)}\Big|^{2}
\end{equation}
where $\sigma(x)$ is an invertible even section of 
$H^{0}(\mathbb{P}^{1|2}, L^{\otimes m})$. 
For $H^{0}(\mathbb{P}^{1|2}, L^{\otimes m}) $ 
we have 
\begin{equation}\label{nuovoJ}
\omega_{m} = \frac{i}{2\pi} \partial\bar\partial \log 
\Big(\sum_{l=0}^{m} |z^{l}_{0} z_{1}^{m-l}|^{2} + i \,
\sum_{l=0}^{m-1} |z^{l}_{0} z_{1}^{m-1-l}|^{2} 
 (\theta_{0} + \theta_{1}) (\bar\theta_{0} + \bar\theta_{1}) - 
\sum_{l=0}^{m-2} |z^{l}_{0} z_{1}^{m-2-l}|^{2} \theta_{1} \bar\theta_{1} \theta_{2} \bar\theta_{2} \Big)\,. 
\end{equation}
The $\theta-\bar\theta$-sections 
are absent in the usual geometry and it appeared in physics 
in the context of supertwistor geometry. The differentials $\partial$ and 
$\bar\partial$ are natural extensions of the one dimensional case. \footnote{
In the case of $H^{0}(\mathbb{P}^{1|2},L^{\otimes m}) $, we can define 
an holomorphic form by separating the commuting  
sections $s_i$ with $i=1, \dots, 2m -1$ from the anticommuting ones 
$\hat s_I$ with $I=1, \dots, 2m$
\begin{equation}\label{nuovoKA}
\Omega_m = \epsilon_{i_1 \dots i_{2m -1}} s_{i_1} d\, s_{i_2} \wedge d\, s_{i_{2m-1}} 
\epsilon_{I_1 \dots I_{2m}} \frac{\partial}{\partial \hat s_{I_1}} \dots 
\frac{\partial}{\partial \hat s_{I_{2m}}}.
\end{equation}}

The next step is to consider a super Hermitian metric 
$L^{\otimes m} \times L^{\otimes m} \rightarrow \mathbb{C}^{1|0}$ defined 
by the formula 
\begin{equation}\label{nuovoK}
h_{m}(q,q') = \frac{1}{\lambda} 
\frac{\frac{q}{\sigma(x)} \overline{\frac{q'}{\sigma(x)}}}{
\sum_{l=0}^{p_{m} + q_{m}}  |s_{l}(x)|^{2}}\,.
\end{equation}
In the denominator, we have both commuting and anticommuting 
sections and they have to be taken into account to define an $L^{2}$-product and an 
orthonormal basis for $\mathbb{P}\Big(H^{0}(M,L^{\otimes m})\Big)$ as follows
\begin{equation}\label{nuovoL}
\langle s_{i}, s_{j} \rangle_{h} = \int_{M} h_{m}(s_{i}(x),s_{j}(x)) 
\Omega_{CY}\wedge \bar\Omega_{CY} = (\delta_{ij}, \epsilon_{ij}) 
\end{equation}
where $\delta_{ij}$ is the diagonal metric for even sections and 
$\epsilon_{ij}$ is the off-diagonal metric for odd sections. The metric $\langle \cdots, \cdots \rangle_{h}$ reduced the symmetry group from $GL(p_{m}|q_{m})$ 
to the supergroup $SU(p_{m}|q_{m})$. 


 
To finish this paragraph, we analyze in detail the balancing of 
$\mathbb{P}^{1|2}$ into $H^0(\mathbb{P}^{1|2},L^{\otimes 2})$, namely into the space of 
homogeneous sections of degree-2. They are given by the set
\begin{equation}\label{exA}
\Big\{ 
z^{a}_{0} z^{2 - a}_{1}, z^{a}_{0} z^{1 -a'}_{1} \theta_I, \theta_1 \theta_2
\Big\}
\end{equation}
where $a=0,1,2$ and $a'=0,1$. 
So, we can form the following integrals (where $z$ is the affine coordinate on 
$\mathbb{P}^1$) 
\begin{eqnarray}\label{exB}
B_{a,b} &=& \int_{\mathbb{P}^{1|2}}  \Omega_{CY} \wedge \bar  \Omega_{CY} 
\frac{z^{2 - a} \bar z^{2 - b}}
{1 + \sum_{a=0,1} |z|^{4-2 a} +  i \sum_{a'=0,1} |z|^{2-2 a'} \theta_I \bar \theta^I + 
\theta_1 \bar \theta_1 \theta_2 \bar \theta_2}\nonumber \\
B_{a' I,b' J} &=& \int_{\mathbb{P}^{1|2}}  \Omega_{CY} \wedge \bar  \Omega_{CY} 
\frac{z^{1 - a'} \bar z^{1 - b'} \theta_I  \bar \theta_J}
{1 + \sum_{a=0,1} |z|^{4-2 a} +  i \sum_{a'=0,1} |z|^{2-2 a'} \theta_I \bar \theta^I + 
\theta_1 \bar \theta_1 \theta_2 \bar \theta_2} \nonumber \\
B_{12, 12} &=& \int_{\mathbb{P}^{1|2}}  \Omega_{CY} \wedge \bar  \Omega_{CY} 
\frac{\theta_1\theta_2 \bar\theta_1\bar\theta_2}
{1 + \sum_{a=0,1} |z|^{4-2 a} +  i \sum_{a'=0,1} |z|^{2-2 a'} \theta_I \bar \theta^I + 
\theta_1 \bar \theta_1 \theta_2 \bar \theta_2} \nonumber \\
\end{eqnarray}
where the integrals are easily performed by the previous instructions. 
In the last integral, the Berezin integration 
removes those $\theta$'s and it leaves a bosonic integral on $\mathbb{P}^1$ which is 
similar to the classical integrals in the bosonic balanced manifolds. However, 
here we see that we have new conditions coming from the other integrals. For example, 
from the first one we need to expand the denominator to 
soak up enough $\theta$'s. And this leads to new conditions on the embeddings. 
Notice that in general the form of the embedding given by (\ref{exA}) is not balanced, 
but one needs to adjust some numerical coefficients in front of each given 
section. 

\subsection{Balancing of points}

As a second application, we consider the problem of the stability 
of point of the type $\mathbb{C}^{0|n}$ into the superprojective space 
$\mathbb{P}^{1|n}$. Since the superspace $\mathbb{C}^{0|n}$ is not a Calabi-Yau, we need 
to use the measure defined in (\ref{thom}). Before doing that, we specify the 
embedding as follows: we denote by $\eta_i, \bar\eta_i$ the anticommuting generators 
of $\mathbb{C}^{0|n}$, we construct the morphism between the two superspace 
by the map
\begin{eqnarray}\label{supA}
&& P:  \mathbb{C}^{0|n}  \longrightarrow \mathbb{P}^{1|n} \nonumber \\
&& X_I = P^*(x_I) =  \alpha_{I} + \alpha_{I,[jk]} \eta_j \eta_k + 
\dots \\
&& \Theta_i = P^*(\theta_i) = \beta_{i,j} \eta_j + \beta_{i,jkl} \eta_j \eta_k \eta_l + \dots 
\end{eqnarray}
where $x_I, \theta_i$ are the homogeneous coordinates on $\mathbb{P}^{1|n}$, with 
$I=0,1$ and $i=1,\dots,n$. The capital letters denote the pull-backs of the 
sheaf generators. 

First we construct the matrices for the bosonic embeddings. For each single point 
$[x_0:x_1]$ the map discussed in (\ref{nuovoA}-\ref{nuovoBB}) gives
\begin{eqnarray}\label{supB}
B{[x_0:x_1]} = \frac{1}{|x_0|^2 + |x_1|^2} 
\left(
\begin{array}{cc}
|x_0|^2 & x_0 \bar x_1 \\
x_1 \bar x_0 & |x_1|^2 
\end{array}
\right) 
\end{eqnarray}
and to extend it to supermanifold, we substitute the pull-backs $X_I$ in place of 
the coordinates $x_I$. In this way, the momentum map becomes a superfield 
of the anticommuting coordinates $\eta_i$ and therefore we need to 
integrate over them to get a numerical value. For that reason we define 
the following new quantity 
 \begin{equation}\label{supC}
M(P) = \int 
\prod_{i=1}^n d\eta_i 
 d\bar\eta_i 
\left(\frac{1}{|X_0|^2 + |X_1|^2 + i \sum_j \Theta_j \bar\Theta_j} 
\left(\begin{array}{cc}
 |X_0|^2 & X_0 \bar X_1 \\
 X_1 \bar X_0 & |X_1|^2 \\
 \end{array}
\right) 
\right)
\end{equation} 
where $P$ is the point in the superprojective space $\mathbb{P}^{1|N}$. 
We also define 
$\sigma_{ij} = \frac{\partial}{\partial \eta_i} \Theta_j(\eta_i)$ as the embedding matrix
\begin{equation}
\sigma_{ij}  = \beta_{i,j} + \beta_{i,j kl} \eta_k \eta_l + \dots 
\end{equation}
The function $M(P)$ is the generalization of the usual 
moment map $B{[x_0:x_1]}$, where we embed the point $[x_0:x_1]$ 
into a $su(2)$ matrix. 
On the other hand, for supermanifolds  $M(P)$ is 
the embedding of the point $P$ into the upper-left corner of the supermatrix 
$su(2|N)$ which is the Lie algebra of the isometry group of $\mathbb{P}^{1|N}$ which is represented by an $su(2)$ matrix. For that reason the normalization term $|X_0|^2 + |X_1|^2 + i \sum_j \Theta_j \bar\Theta_j$ acquires the supplementary summand $i \sum_j \Theta_j \bar\Theta_j$. 
Notice that we have also to  take into account the embedding of the point into 
the $su(N)$ part of the supermatrix needed to implement the third type of condition in eqs. (\ref{nuovoE}) and this will be done later.

So, the final condition for the stability of 
a set of points $\mathbb{C}^{0|n}$ immersed into $\mathbb{P}^{1|n}$ is 
\begin{eqnarray}\label{subD}
&&\sum_P 
\prod_{i=1}^n 
\frac{\partial}{\partial \eta_i} 
\frac{\partial}{\partial \bar\eta_i} 
\left. 
{\mathcal M}_P
\right|_{\eta_i = \bar \eta_i=0} = \lambda {\mathbf 1} \,, \nonumber \\
&&\vspace{4cm}
{\mathcal M}_P = 
\frac{1}{|X_{P,0}|^2 + |X_{P,1}|^2  + i \sum_j \Theta_{P,j} \bar\Theta_{P,j}} 
\left(\begin{array}{cc}
 |X_{P,0}|^2 & X_{P,0} \bar X_{P,1} \\
 X_{P,1} \bar X_{P,0} & |X_{P,1}|^2 \\
 \end{array}
\right)
\end{eqnarray} 
where $X_{P,I}, \Theta_{P,i}$ are respectively the embeddings 
for the point P and the sum is extended over all points. 

Before discussing the third type of condition in eqs. (\ref{nuovoE}), let us analyze the condition (\ref{subD}) for a specific example, when $\mathbb{C}^{0|2}$ 
is embedded into 
$\mathbb{P}^{1|2}$. 
For that we consider the embedding
\begin{eqnarray}\label{exaA}
&& P:  \mathbb{C}^{0|2}  \longrightarrow \mathbb{P}^{1|2} \nonumber \\
&& X_i = P^*(x_i) =  \alpha_{i} + \tilde\alpha_{i} \eta_1 \eta_2 \\
&& \Theta_i = P^*(\theta_i) = \sigma_{i,j} \eta_j \nonumber
\end{eqnarray}
and we compute explicitly the expression in (\ref{subD}). 
After few manipulations, we get
\begin{eqnarray}\label{exaB}
&&\sum_P 
\left\{ 
\Upsilon^3_P \left( 2 |\sum_i  \bar\alpha_{P,i} \tilde\alpha_{P,i} |^2 - 
\sum_i  |\alpha_{P,i} |^2 \sum_i  |\tilde\alpha_{P,i} |^2  - 2 \, (\det|\sigma|^2)
\right) \left(\begin{array}{cc} |\alpha_{P,0}|^2 & \alpha_{P,0} \bar\alpha_{P,1} \\
 \alpha_{P,1} \bar\alpha_{P,0} & |\alpha_{P,1}|^2
\end{array}\right) \nonumber \right.\\
&&
+\Upsilon^2_P \left[ 
\left( \sum_i \alpha_{P,i} \tilde{\bar\alpha}_{P,i} \right)
\left(\begin{array}{cc} \alpha_{P,0} \tilde{\bar\alpha}_{P,0} & \alpha_{P,1} \tilde{\bar\alpha}_{P,0} \\
 \alpha_{P,0} \tilde{\bar\alpha}_{P,1} & \alpha_{P,1} \tilde{\bar\alpha}_{P,1}
\end{array}\right) + {\rm h.c.} \right]   \nonumber \\
&&
-\left. \Upsilon_P 
\left(\begin{array}{cc} |\tilde\alpha_{P,0}|^2 & \tilde\alpha_{P,0} \tilde{\bar\alpha}_{P,1} \\
 \tilde\alpha_{P,1} \tilde{\bar\alpha}_{P,0} & |\tilde\alpha_{P,1}|^2
\end{array}\right) 
\right\} = \lambda {\mathbf 1}\,.
\end{eqnarray}
where $\Upsilon_P = 1/ \sum_i |\alpha_{P,i}|^2$.
For example, in the case of only a single point $P = [1:0]$, we 
get the simplified equation
\begin{eqnarray}\label{exaC}
\left( 
\begin{array}{cc}
- |\tilde\alpha_1|^2 + \tilde\alpha_0^2 + \tilde{\bar\alpha}_0^2 - 
2 \det|\sigma|^2
& \tilde{\alpha}_0 \tilde{\alpha}_1 - 
\tilde{\alpha}_0 \tilde{\bar\alpha}_1 \\
\tilde{\bar\alpha}_0 \tilde{\bar\alpha}_1 - \tilde{\bar\alpha}_0 \tilde{\alpha}_1& 
- |\tilde\alpha_1|^2
\end{array}
\right) = \lambda {\mathbf 1}\,. 
\end{eqnarray}
From the up-right corner we get $ \tilde\alpha_1$ is real and this fixes 
the constant $\lambda$. Then, we get the condition, $\tilde\alpha_0^2 + \tilde{\bar\alpha}_0^2 - 2\, \det|\sigma|^2 
= 0$ which can be solved in terms of $\tilde\alpha_0$. 
Therefore, there is a single point whose embedding 
into $\mathbb{P}^{1|2}$ is balanced. The logic can be repeated for several 
points and other solutions can be also found. Notice that the non-numerical part of 
$X_i$, namely the part which is parametrized by $\tilde\alpha_i$ plays a fundamental 
role and serves for the balancing. We can also recover the classical solution by setting all 
$\tilde\alpha_i$ to zero. This implies the classical balancing condition and therefore we found that 
there are also the classical solutions with anticommuting coordinates.

It remains to compute the contribution for the embedding in the second $su(2)$ of the supermatrix $su(2|2)$ and 
for that we have 
 \begin{equation}
 M(P)_{k l} = \int 
\prod_{i=1}^n d\eta_i 
 d\bar\eta_i 
\left(\frac{i \Theta_k \bar\Theta_l}{|X_0|^2 + |X_1|^2 + i \sum_j \Theta_j \bar\Theta_j} 
\right)
\end{equation}
as follows from (\ref{nuovoE}). It is easy to evaluate the Berezin integrals to get
\begin{equation}\label{supCA}
M(P)_{k l}  = \lambda \delta_{k l}
\end{equation}
where $\lambda = \det|\sigma|^2/ (\sum_i |\alpha_i|^2)^2$ and in the case of the point $P=[1:0]$ 
we have $\lambda =\det|\sigma|^2$. 
In the way, we notice that this part 
of the embedding is automatically balanced and it does not yield a new condition on the 
parameters of the embedding. 

To our knowledge, the present discussion is a way to formulate the balancing of points 
into a superprojective space. Of course, one can add further condition, for example inserting in the 
integral the factor $\exp ( i \sum_j \Theta_j \bar\Theta_j)$. This term reproduces the previous results, but 
in addition it leads to a further condition that coincides with the classical requirement of balanced points into a 
projective space. So, instead of imposing  by hands the additional condition of classical stability, the 
modification of the integration measure yields all possible set of conditions. Moreover, for 
the case with more than 2 anticommuting coordinates the exponential factor  $\exp (i  \sum_j \Theta_j \bar\Theta_j)$ 
will lead to new conditions on the embedding. This makes sense since adding new anticommuting 
coordinates requires new embedding parameters to be fixed. 

There are several questions that can be addressed in the same framework, for example: 
can all points be made balanced after an $SL(2|2)$ transformation? Is there a relation between our definition of balanced supermanifolds and a suitable notion of stability, such as in GIT, in the supermanifold context (see \cite{thomas})? We will leave 
these questions to forthcoming publications.

\section*{Acknowledgments}
We thank P. Aschieri, U. Bruzzo, L. Castellani, P. Fr\'e, and E. Scheidegger for useful discussions and 
remarks.


\end{document}